\pdfoutput=1

\documentclass[11pt,nofootinbib]{article}
\usepackage{amsmath,amssymb,mathtools}
\usepackage{graphicx}
\usepackage[colorlinks=true,linkcolor=blue,citecolor=blue,urlcolor=blue,linktocpage=true]{hyperref}
\usepackage{subfigure}
\usepackage{comment}
\usepackage{tensor}
\usepackage{cases}

\numberwithin{equation}{section}

\begin{document}

\providecommand{\abs}[1]{\lvert#1\rvert}
\providecommand{\bd}[1]{\boldsymbol{#1}}

\begin{titlepage}

\setcounter{page}{1} \baselineskip=15.5pt \thispagestyle{empty}

\begin{flushright}
SISSA 39/2016/FISI,~TU-1026,~IPMU16-0118
\end{flushright}
\vfil

%\vspace{2.5cm}

\bigskip
\begin{center}
 {\LARGE \textbf{Cosmological Perturbations of}}\\
\medskip 
 {\LARGE \textbf{Axion with a Dynamical Decay Constant}}
\vskip 15pt
\end{center}

\vspace{0.5cm}
\begin{center}
{\Large 
Takeshi Kobayashi$^{\star, \dagger}$
and
Fuminobu Takahashi$^{\ast, \ddagger}$
}\end{center}

\vspace{0.3cm}

\begin{center}
\textit{$^{\star}$ SISSA, Via Bonomea 265, 34136 Trieste, Italy}\\

\vskip 14pt
\textit{$^{\dagger}$ INFN, Sezione di Trieste, Via Bonomea 265,
 34136 Trieste, Italy}\\ 
 
\vskip 14pt
 \textit{$^{\ast}$ Department of Physics, Tohoku University, Sendai,
 Miyagi 980-8578, Japan}\\

\vskip 14pt
 \textit{$^{\ddagger}$ Kavli IPMU (WPI), UTIAS, The University of Tokyo,
 Kashiwa, Chiba 277-8583, Japan}\\ 

\vskip 14pt
E-mail: 
\texttt{\href{mailto:takeshi.kobayashi@sissa.it}{takeshi.kobayashi@sissa.it}}, 
\texttt{\href{mailto:fumi@tuhep.phys.tohoku.ac.jp}{fumi@tuhep.phys.tohoku.ac.jp}}

\end{center} 

%\vfil

%\vspace{0.8cm}

\vspace{1cm}

\noindent
A QCD axion with a time-dependent decay constant has been known to be
 able to accommodate high-scale inflation without producing topological
 defects or too large isocurvature perturbations on CMB scales.
 We point out that a dynamical decay constant also has the effect of
 enhancing the small-scale axion isocurvature perturbations.
 The enhanced axion perturbations can even exceed the periodicity of the
 axion potential, and thus lead to the formation of axionic domain
 walls. Unlike the well-studied axionic walls, the walls produced from the
 enhanced perturbations are not bounded by cosmic strings, and thus
 would overclose the universe independently of the number of degenerate
 vacua along the axion potential.
 
\vfil

\end{titlepage}

\newpage
\tableofcontents

\section{Introduction}
\label{sec:intro}

The QCD axion is a Nambu--Goldstone boson which arises 
in association with spontaneous breakdown of Peccei--Quinn (PQ) symmetry,
and it dynamically solves the strong CP problem~\cite{Peccei:1977hh,Weinberg:1977ma,Wilczek:1977pj}.
The strength of interactions with the standard model particles as well as the periodicity of the
axion potential are determined by the axion decay constant, which is
currently constrained as 
\begin{equation}
 10^9\, \mathrm{GeV} \lesssim f_a \lesssim 10^{12}\, \mathrm{GeV}.
  \label{f-bound}
\end{equation}
Here the lower bound is set by astrophysical arguments of star
cooling~\cite{Raffelt:1996wa}, while
the upper bound comes from the requirement that the abundance of the
axion be less than or equal to that of cold dark matter (CDM) without
fine-tuning the initial misalignment angle~\cite{Preskill:1982cy,Abbott:1982af,Dine:1982ah}.

Cosmological considerations further set bounds on a combination of the decay
constant~$f_a$ and the inflationary Hubble scale~$H_{\mathrm{inf}}$.
Firstly, $f_a$ should be larger than $H_{\mathrm{inf}}$, otherwise the PQ
symmetry breaking would happen after inflation and thus would produce
domain walls which overclose the universe.
An exception to this statement is when the number of degenerate vacua
along the bottom of the PQ scalar's Mexican hat potential (the so-called
domain wall number) is $N = 1$;
then the walls are bounded by cosmic strings
without being connected to other walls, and such walls of finite
size can annihilate soon after the QCD phase transition~\cite{Vilenkin:1982ks,Sikivie:1982qv,Linde:1990yj,Lyth:1992tx}. 

Furthermore, if the axion constitutes the CDM,
then even tighter bounds on $f_a$ and $H_{\mathrm{inf}}$ 
are obtained from constraints on CDM isocurvature
perturbations~\cite{Seckel:1985tj,Lyth:1989pb}.
For instance, with a decay constant of $f_a = 10^{12}\, \mathrm{GeV}$,
current constraints on CDM isocurvature
from measurements of the cosmic microwave background
(CMB)~\cite{Ade:2015lrj} require  
$H_{\mathrm{inf}} \lesssim 10^7\, \mathrm{GeV}$.
The bound on~$H_{\mathrm{inf}}$ is particularly strong
for $f_a \lesssim 10^{11}\, \mathrm{GeV}$,
due to the nonquadratic form of the axion potential~\cite{Lyth:1991ub},
dubbed the anharmonic effect. 
The anharmonic enhancement of the axion isocurvature perturbations
becomes so strong for small values of~$f_a$ 
that a decay constant of
$f_a \lesssim 10^9\, \mathrm{GeV}$ is excluded~\cite{Kobayashi:2013nva};
thus the lower bound of~(\ref{f-bound}) can also be derived from
cosmology when the dark matter consists of axions.
These observations indicate that, any detection of primordial
gravitational waves from inflation in the near future would exclude the
QCD axion as a dark matter candidate. 

Here it should be noted that
the above arguments on $f_a$ and $H_{\mathrm{inf}}$ from
domain walls and isocurvature perturbations are modified
if the decay constant takes different values
during the inflation epoch and today.
The works~\cite{Linde:1990yj,Linde:1991km} pointed out that
if $f_a$ was larger during inflation,
then the axion isocurvature
perturbations would actually be smaller than that inferred from the
present value of~$f_a$, allowing high-scale inflation without
contradicting observations.
(See also
Refs.~\cite{Higaki:2014ooa,Chun:2014xva,Fairbairn:2014zta,Kearney:2016vqw}
for recent works along this line.
A similar effect can also be induced by a non-minimal coupling 
of the axion kinetic term to gravity~\cite{Folkerts:2013tua}.)
While this is certainly an intriguing possibility, however,
the introduction of a dynamical decay constant not only
leads to smaller-than-expected isocurvature perturbations on the CMB scales,
but also {\it dynamically enhances} the isocurvature perturbations on smaller length
scales. This is understood by viewing the decay constant~$f_a$ and the axion
field~$\theta$ respectively as the radial and angular directions of the
PQ scalar;
if the radial component~$f_a$ is forced to quickly shrink while the
angular velocity~$\dot{\theta}$ is nonzero, then the motion along the
angular direction would speed up, leading to the growth of the
fluctuation~$\delta \theta$ among different regions of space.
Moreover, this effect not only works on sub-horizon scales,
but further stretches out to scales much larger than
the Hubble radius at the time when the decay constant varies.\footnote{A
similar effect was invoked in~\cite{Kobayashi:2014sga} for the generation of
large-scale magnetic fields in the post-inflationary universe.}

In this paper we examine the evolution of the full axion perturbation
spectrum under a time-dependent decay constant. 
We show that, while the dynamical decay constant can help to evade the isocurvature
constraints on CMB scales,
it can also strongly enhance the axion fluctuations on smaller scales,
which may even exceed the periodicity of the axion potential and thus 
lead to the formation of axionic domain walls.
These domain walls are produced due to the enhanced axion fluctuations,
although the PQ symmetry is broken before inflation,
and thus they are different from the well-studied axionic walls.
In particular, the axionic walls from the enhanced fluctuations
are not bounded by cosmic strings, and thus would overclose the universe
even in the case of $N = 1$~\cite{Preskill:1991kd}. 
The strong enhancement of the axion fluctuations may also lead to the
recovery of the PQ symmetry in the post-inflation epoch, which would
upset the cosmological expansion history.\footnote{In this paper we investigate the enhancement of the sub- and super-horizon scale axion fluctuations due to the shrinking of the decay constant. 
We also note that if the radial component of the PQ scalar oscillates
about its minimum, both the axion and the radial component can be
produced by parametric resonance. Such sub-horizon fluctuations can 
lead to a nonthermal symmetry restoration, which has been studied
in~\cite{Tkachev:1995md,Kasuya:1996ns,Kasuya:1997ha,Kasuya:1998td,Tkachev:1998dc,Kearney:2016vqw}.}
On the other hand, if the axion perturbations are only mildly enhanced,
then such small-scale isocurvature could cause a variety of observable
consequences, as discussed in
e.g.~\cite{Chluba:2013dna,Sekiguchi:2013lma}.

Before moving on, we also comment on another possibility to suppress the axionic isocurvature
perturbations. If the PQ symmetry is badly broken during inflation, the axion mass may
become heavier than or comparable to the Hubble parameter. Then,
it does not acquire sizable super-horizon fluctuations, suppressing
the isocurvature 
perturbations~\cite{Dine:2004cq,Higaki:2014ooa,Dine:2014gba,Kawasaki:2015lea,Takahashi:2015waa,Kawasaki:2015lpf,Nomura:2015xil}. In this case, one has to make sure that the explicit 
PQ symmetry breaking is sufficiently suppressed in the present universe to be consistent 
with the neutron electric dipole moment experiments.

The layout of this paper is as follows:
We describe the setup and notation in Section~\ref{sec:setup}.
In Section~\ref{sec:dynamics} we give general discussions on the
evolution of the axion 
fluctuations under a dynamical decay constant,
then in Section~\ref{sec:DDC} we 
compute the fluctuations in the post-inflationary universe. 
In Section~\ref{sec:cosmo}, we discuss the cosmological implications of
the enhanced fluctuations, and also carry out case studies.
We conclude in Section~\ref{sec:conc}.

\section{QCD Axion with a Dynamical Decay Constant}
\label{sec:setup}

Throughout this paper we denote the PQ symmetry breaking scale by~$f$, i.e.,
the field value of the PQ scalar at the minimum of its Mexican hat
potential being $\abs{\Phi}_{\mathrm{min}} = f / \sqrt{2}$. 
Writing the PQ scalar in terms of two real scalar fields $r$ and $\theta$ as
\begin{equation}
 \Phi = \frac{f + r }{\sqrt{2}}e^{i \theta },
\end{equation}
then the angular direction~$\theta$ along the potential minimum serves as the
axion field. We  refer to $f$  and $\theta$ as the axion decay constant
and the (dimensionless) axion field, respectively.

We couple $\Phi$ to $N$ heavy PQ quark pairs $(Q_i, \bar{Q}_i)$ in the fundamental
and anti-fundamental representation of SU(3)$_c$, where the flavor index $i$ runs from $1$ to $N$. 
 Then, the axion couples to gluons through the
color anomaly of the PQ symmetry. Integrating out the heavy PQ quarks and setting the radial component
to the potential minimum, the action of the axion is given by
\begin{equation}
 S = \int d^4 x \sqrt{-g}
  \left\{
-\frac{1}{2} f^2 g^{\mu \nu} \partial_\mu \theta \partial_\nu \theta
 + N \theta \frac{g_s^2}{32 \pi^2}G_{A \mu \nu} \tilde{G}_A^{\mu \nu}
	 \right\}.
\end{equation}
This gives an effective potential for $\theta$ with periodicity $2 \pi / N$,
\begin{equation}
S = \int d^4 x \sqrt{-g}
 \left\{
-\frac{1}{2} f^2 g^{\mu \nu} \partial_\mu \theta \partial_\nu \theta
 - m_{\mathrm{QCD}}^2(T) \frac{f^2}{N^2} \left(
 1 - \cos \left(N \theta \right)
 \right)
	  \right\}.
 \label{theta-action}
\end{equation}
Here, the temperature-dependent axion mass~$m_{\mathrm{QCD}}(T)$ emerges when the
cosmic temperature 
cools down to around $\Lambda_{\mathrm{QCD}} \approx 200\, \mathrm{MeV}$. 
On the other hand at $T \gg \Lambda_{\mathrm{QCD}}$,
the axion is a free field with only the kinetic term.
We also note that the discussions in Section~\ref{sec:intro} can be
rewritten for~$f$ by the replacement $f_a \to f/N$; in particular, 
(\ref{f-bound}) is read as a bound on $f/N$.

Now, as we have discussed in the introduction,
the coefficient of the kinetic term can be time-dependent when the decay
constant~$f$ is actually a dynamical field~\cite{Linde:1990yj,Linde:1991km}.
(See also e.g.~\cite{Kearney:2016vqw,Kasuya:1996ns,Kasuya:2009up} for
explicit models of axions with dynamical decay constants.)
Let us now study the axion dynamics with a dynamical~$f$,
at very early times when the axion potential is absent.
We consider a flat FRW background
\begin{equation}
 ds^2 = a(\tau)^2 ( -d\tau^2 + d \bd{x}^2),
\end{equation}
and assume the decay constant to be homogeneous, i.e. $f =
f(\tau)$. Then the equation of motion of the axion reads
\begin{equation}
 \frac{(a^2 f^2 \theta')'}{a^2 f^2} - \delta^{ij} \partial_i \partial_j
  \theta = 0, 
\end{equation}
where a prime denotes a derivative in terms of the conformal
time~$\tau$. Going to the Fourier space,
\begin{equation}
 \theta (\tau, \bd{x}) = \frac{1}{(2 \pi)^3}\int d^3 k \, 
  e^{i \bd{k}\cdot\bd{x}} \theta_{\bd{k}}(\tau),
\end{equation}
the equation of motion is rewritten as
\begin{equation}
  \frac{(a^2 f^2 \theta_{\bd{k}}')'}{a^2 f^2} + k^2 \theta_{\bd{k}} =
   0.
   \label{EOM2}
\end{equation}
We stress that this is the exact equation of motion of the axion before
the potential emerges, and in particular that there is no mode-mode coupling. 

When quantizing the theory by promoting $\theta$ to an operator, one
can check that the commutation relations of $\theta$ and its conjugate
momentum are equivalent to those of the annihilation and creation
operators when the mode functions are independent of the direction
of~$\bd{k}$, i.e.,
\begin{equation}
 \theta_{\bd{k}} = \theta_k
\end{equation}
with $k = \abs{\bd{k}}$, and further obey the normalization condition
\begin{equation}
 \theta_k \theta_k'^* - \theta_k^* \theta_k' = \frac{i}{a^2 f^2}.
  \label{norm}
\end{equation}
By choosing the vacuum $| 0 \rangle$ to be annihilated by
all the annihilation operators, the
two-point correlation function of the 
axion field is computed as 
\begin{equation}
 \langle \theta (\tau, \bd{x}) \, \theta (\tau, \bd{y}) \rangle
  = \int \frac{d^3 k}{4 \pi k^3}
  e^{i \bd{k} \cdot (\bd{x} - \bd{y})}
  \mathcal{P}_\theta (\tau, k),
\end{equation}
where the power spectrum is
\begin{equation}
 \mathcal{P}_\theta (\tau, k) = \frac{k^3 }{2 \pi^2 }
  \left| \theta_k (\tau) \right|^2.
  \label{Poftheta}
\end{equation}

\section{Dynamics of Axion Perturbations}
\label{sec:dynamics}

Although the axion potential is absent in the very early universe,
the axion velocity, i.e. the angular velocity of the PQ scalar,
is not exactly zero due to quantum fluctuations,
which are stretched to cosmic scales and become classical during
inflation.
The velocity fluctuations can get strongly enhanced even on
super-horizon scales 
when the decay constant~$f$, corresponding to the radial component of
the PQ scalar, quickly shrinks.
The enhanced velocity fluctuations would then lead to the growth of the axion
field fluctuations. 

The basic picture can be seen by looking at the homogeneous mode
$k = 0$, for which the 
equation of motion~(\ref{EOM2}) gives the scaling of
\begin{equation}
 \theta_0' \propto \frac{1}{a^2 f^2 }.
  \label{0scaling}
\end{equation}
This clearly shows that the phase velocity, given an initially nonzero value,
increases when $f$ decreases quickly enough.
However we should also note that for a nonzero $k$ mode, even when it is
well outside the horizon, the
scaling behavior of the velocity fluctuation is not necessarily the same
as for the homogeneous mode~(\ref{0scaling}).

Let us now look into the inhomogeneous modes, i.e. $k \neq 0 $,
which represent the axion fluctuations.
To make the discussions concrete, we focus on an FRW background with a
constant equation of state parameter~$w$ that is not $-1/3$,
i.e.,
\begin{equation}
 \frac{p_{\mathrm{bg}}}{\rho_{\mathrm{bg}}} = w =
  \mathrm{const.} \neq - \frac{1}{3}.
\end{equation}
Then for the Hubble rate $H = a' / a^2$, its time derivative obeys
\begin{equation}
 \frac{H'}{a H^2} = -\frac{3 (1+w)}{2}.
\end{equation}
The condition of $w \neq -1/3$ guarantees that $ a H $ does not stay
constant, which turns out to be convenient when we rewrite the equation of
motion below. 
We further assume that the decay constant varies in time as a power-law
of the scale factor,
\begin{equation}
 f \propto a^{-n},
\end{equation}
with a constant~$n$.
Then, introducing
\begin{equation}
 u \equiv \frac{2 }{\abs{1 + 3 w }}\frac{k}{a H},
  \qquad
  \nu \equiv \frac{1}{2} - \frac{2 (1-n)}{1 + 3 w },
\end{equation}
the equation of motion~(\ref{EOM2}) can be rewritten in the form
\begin{equation}
 \frac{d^2 \theta_k}{d u^2} + \frac{1 - 2 \nu }{u}
  \frac{d \theta_k}{du } + \theta_k = 0.
\label{EOMwith_u}
\end{equation}
The general solution of this equation is
\begin{equation}
 \theta_k = u^\nu \left\{
 \kappa  Z_{\nu}^{(1)} (u) + \lambda  Z_{\nu}^{(2)} (u)
	   \right\},
 \label{generalsol}
\end{equation}
where 
($Z_{\nu}^{(1)}$, $Z_{\nu}^{(2)}$) is a pair of
Hankel or Bessel functions of the first and second kinds,
and $\kappa$, $\lambda$ are constants.
The phase velocity can be expressed
in terms of lower order functions as
\begin{equation}
 \frac{d \theta_k }{d\ln a}= \frac{1 + 3 w }{2} u^{\nu + 1}
 \left\{
  \kappa  Z_{\nu - 1}^{(1)} (u) + \lambda  Z_{\nu - 1}^{(2)} (u)
	      \right\} ,
 \label{gendthetada}
\end{equation}
where we have used a relation for the Hankel/Bessel functions,
\begin{equation}
\frac{d Z_\nu(u)}{du} = Z_{\nu-1}(u)-\frac{\nu}{u} Z_\nu(u).
\end{equation}
We also note that the variable $u$ varies in time as $u \propto a^{(1+3w)/2}$ in
terms of the scale factor, 
and $u \to 0 $ $(u\to \infty)$ describes the modes being well outside (inside) the Hubble horizon.

\subsection{Outside the Horizon}

Let us study the behavior of the solution (\ref{generalsol})  when 
the modes are well outside the horizon.
Here we discuss the general solution using
Bessel functions, i.e.,
\begin{equation}
 Z_\nu^{(1)} (u) = J_{\nu}(u),
  \qquad
Z_\nu^{(2)} (u) = Y_{\nu}(u).
\end{equation}
The limiting forms of the Bessel functions for $u
\to 0$ are~\cite{Olver:2010:NHMF}, 
\begin{equation}
 \begin{split}
J_{\nu - 1}(u) \sim \frac{1}{\Gamma (\nu ) }
 \left( \frac{u}{2} \right)^{\nu        
 - 1},
    & 
  \qquad
 Y_{\nu - 1}(u) \sim - \frac{\Gamma (\nu - 1 ) }{\pi}
  \left( \frac{u}{2} \right)^{1
 -\nu},
 \\
  & \mathrm{when}\; \nu > 1 \;  \mathrm{or}  \;  \nu = 1/2, \, -1/2,
 \,  -3/2,  \, \cdots.
 \end{split}\label{Besseluto0}
\end{equation}
Here, since the leading order expressions of 
$J_{\nu-1}$ and $Y_{\nu-1}$ scale differently in term
of~$u$,
they do not continually cancel each other in
the expression~(\ref{gendthetada}) for the phase velocity.\footnote{On
the other hand for  
$\nu < 1 $, $ \nu \neq 1/2, \, -1/2,  \, 
-3/2,  \, \cdots$, the 
two kinds of the Bessel function at $u \to 0$ can scale in the same way 
as $J_{\nu - 1} \propto Y_{\nu - 1} \propto u^{\nu - 1}$.
In such cases the $\kappa$ and $\lambda$ terms can continually cancel each other at
the leading order and so the analyses become a bit tricky.}
Thus on super-horizon scales $ k \ll a H$, the $\kappa$ and $\lambda$
terms in~(\ref{gendthetada}), together with the
prefactor $u^{\nu + 1}$, scale as
$\propto a^{\nu (1 + 3 w)} $ and $ \propto a^{1 + 3 w}$,
respectively; whichever grows faster eventually dominates the
super-horizon phase velocity.

To summarize, unless one of the constants 
$\kappa$, $\lambda$ are set to zero, the phase velocity on modes well
outside the horizon $k \ll aH$ scales eventually as 
\begin{numcases}{ \frac{d \theta_k}{d\ln a} \propto }
 a^{\nu (1 + 3 w) } & \negthickspace \negthickspace \negthickspace for
 ($w > -1/3, \;  \nu > 1$)
 or
 ($w < -1/3, \; \nu = 1/2, \, -1/2, \, -3/2, \cdots$),
 \label{case-ab}
 \\
 a^{1 + 3 w } & \negthickspace \negthickspace \negthickspace for
 ($w < -1/3, \; \nu > 1$)
 or
 ($w > -1/3, \; \nu = 1/2, \, -1/2, \, -3/2, \cdots$).
 \label{case-cd}
\end{numcases}
The latter case~(\ref{case-cd}) is not captured by the
analysis of the homogeneous mode~(\ref{0scaling}). 
One can already see in the equation of motion~(\ref{EOMwith_u}) that 
the super-horizon limit of the inhomogeneous modes may
behave differently from an exactly homogeneous mode;
there $k = 0$ corresponds to $u = 0$, and thus the dynamics 
can be essentially different between homogeneous and inhomogeneous modes. 
We also note that, although the parameter ranges for $(w, \nu)$ shown in
(\ref{case-ab}) and (\ref{case-cd}) do not cover all the possibilities,
they will be sufficient for the cases we discuss in the following sections. 

Expressing the scaling behaviors of the two cases
(\ref{case-ab}) and (\ref{case-cd}) collectively as
\begin{equation}
 \frac{d \theta_k}{d \ln a }\propto a^{p},
  \label{dthetada-p}
\end{equation}
with $p$ being either $\nu (1+3 w)$ or $(1 + 3 w)$,
then one sees that the phase velocity corresponds to a dynamical mode for
the phase fluctuations which scales as
\begin{equation}
 \theta_k^{\mathrm{dyn}} \propto a^p.
\end{equation}
The total super-horizon phase fluctuation contains this dynamical
mode as well as the usual constant mode.

\subsection{Inside the Horizon}

The behavior of the modes when deep inside the horizon is seen from the
asymptotic forms of the Bessel functions for $ u \to \infty$
(recall that $u$ is a positive variable),
\begin{equation}
 J_{\nu} (u)  \sim \sqrt{ \frac{2 }{\pi u } }
 \cos \left( u - \frac{\nu \pi }{2} - \frac{ \pi  }{4} \right),
 \qquad
 Y_{\nu} (u)  \sim \sqrt{ \frac{2 }{\pi u } }
 \sin \left( u - \frac{\nu \pi }{2} - \frac{ \pi  }{4} \right).
\end{equation}
Therefore, $\theta_k$ inside the horizon is an oscillating solution
with an oscillation amplitude~$\tilde{\theta}_k$ that scales in time as
\begin{equation}
 \tilde{\theta}_k \propto u^{\nu - \frac{1}{2}}  \propto a^{n-1} .
  \label{inside-hori}
\end{equation}
In particular when~$f$ is a constant, i.e. $n = 0$, the fluctuation
simply damps as $\propto a^{-1}$ due to the expansion of the universe.

\section{Evolution of Axion Perturbations after Inflation}
\label{sec:DDC}

Let us now apply the above discussions to study the evolution of axion
perturbations under a dynamical~$f$ in the post-inflation epoch.
We suppose the PQ symmetry to be broken before inflation,
and we particularly study the case where $f$ takes a constant
value~$f_{\mathrm{inf}}$ until 
some time after inflation,\footnote{A case where $f$ varies during
inflation was discussed in~\cite{Kasuya:2009up}, in order to generate a
blue-tilted isocurvature spectrum.} 
then varies with time as $\propto a^{-n}$
until it reaches the present-day value~$f_0$, i.e.,
\begin{equation}
  f = 
 \begin{dcases}
     f_{\mathrm{inf}}
       & \text{for $a \leq a_\mathrm{i}$,} \\
     f_{\mathrm{inf}} \biggl( \frac{a_\mathrm{i}}{a} \biggr)^{n}
       & \text{for $a_\mathrm{i} <  a \leq a_\mathrm{f}$,} \\
     f_{\mathrm{inf}} \biggl( \frac{a_\mathrm{i}}{a_\mathrm{f}} \biggr)^{n} = f_0 
       & \text{for $a > a_\mathrm{f}$.}
 \end{dcases}
 \label{f-evo}
\end{equation}
Here $a_\mathrm{i}$ and $a_\mathrm{f}$ denote the scale factors in the
post-inflation epoch when the time-evolution of~$f$ starts and terminates,
respectively.
Quantities measured at $a = a_{\mathrm{i(f)}}$ will be represented by
the subscripts~``$\mathrm{i(f)}$''. 
We further assume that the universe becomes (effectively)
matter-dominated, i.e. $w = 0$, right after inflation, and that $f$
evolves in time during 
this phase.

The dynamics of the axion fluctuations can be
analyzed by connecting the solutions for $\theta_k$ in each epoch.
However for clarity, instead of connecting the exact solutions
(\ref{generalsol}), (\ref{gendthetada}),
we will mostly use the asymptotic scaling solutions we have obtained for the
super-horizon modes~(\ref{case-ab}), (\ref{case-cd}), and the
sub-horizon modes~(\ref{inside-hori}).
This will provide useful approximations, as we will see later when we compare
with the exact solutions.

\subsection{Inflation Epoch}

The axion fluctuations during inflation with
constant Hubble rate $H = H_{\mathrm{inf}}$
and decay constant $f = f_{\mathrm{inf}}$
is given by (\ref{generalsol}) with $w = -1$ and $\nu = 3/2$.
The coefficients $\kappa$ and $\lambda$ are fixed from the normalization
condition~(\ref{norm}) and the requirement that the mode function
approaches a positive frequency solution in the asymptotic past,
i.e., requiring a Bunch--Davies vacuum.
For this purpose it is convenient to use the Hankel function of the
first kind, with which the solution is expressed as,
up to an unimportant phase factor,
\begin{gather}
 \theta_k = \frac{\pi^{1/2}}{2 a^{3/2} 
  H_{\mathrm{inf}}^{1/2} f_{\mathrm{inf}} }
 H^{(1)}_{3/2} \left(\frac{k}{a H_{\mathrm{inf}}}\right),
 \label{4.2}
\\
 \frac{d \theta_k}{d \ln a} = - \frac{\pi^{1/2} k }{2 a^{5/2} 
   H_{\mathrm{inf}}^{3/2} f_{\mathrm{inf}}}
 H^{(1)}_{1/2} \left(\frac{k}{a H_{\mathrm{inf}}}\right).
 \label{4.3}
\end{gather} 
In the super-horizon limit $ k \ll a H_{\mathrm{inf}}$, these
solutions are approximated by 
\begin{align}
 \theta_k &\sim -\frac{i}{2^{1/2}}
 \frac{H_{\mathrm{inf}}}{k^{3/2} f_{\mathrm{inf}} }
  \equiv \theta_k^{\mathrm{const}},
 \label{inf-theta-sh}
\\
 \frac{d\theta_k}{d \ln a} & \sim \frac{i}{2^{1/2}}
 \frac{k^{1/2}}{a^2  H_{\mathrm{inf}} f_{\mathrm{inf}} }
 = - \theta_k^{\mathrm{const}} 
 \left( \frac{k}{a H_{\mathrm{inf}}} \right)^2 .
 \label{inf-dthetada-sh}
\end{align}
One clearly sees from (\ref{inf-theta-sh}) that $\theta_k$ is constant
at leading order (we denote this value by $\theta_k^{\mathrm{const}}$ hereafter),
and that the dynamical mode described by the phase
velocity~(\ref{inf-dthetada-sh}) is subdominant on super-horizon
scales.

\subsection{Post-Inflation Epoch with Constant~$f$}

The above solutions provide the initial conditions for
the axion fluctuations in the subsequent matter-dominated era with a
constant~$f$. 
Focusing on super-horizon modes, i.e. $k \ll a H $,
let us obtain the phase velocity by connecting
(\ref{inf-dthetada-sh}) at the end of inflation $a= a_{\mathrm{end}}$
to the solution~(\ref{case-cd}) with $w = 0$ (note that $\nu = -3/2$ during this
epoch), giving
\begin{equation}
 \frac{d\theta_k}{d \ln a} \sim
  -\theta_k^{\mathrm{const}} \left( \frac{k}{a_{\mathrm{end}}
			      H_{\mathrm{inf}}} \right)^2 
  \frac{a}{a_{\mathrm{end}}} 
  =
  -\theta_k^{\mathrm{const}} \left( \frac{k}{a H} \right)^2.
  \label{I-dthetada-sh}
\end{equation}
It is clear that the dynamical mode obtained by integrating
(\ref{I-dthetada-sh}) is subdominant compared to the constant
mode inherited from the inflation epoch~(\ref{inf-theta-sh});
hence the fluctuation at the leading order is still
\begin{equation}
 \theta_k \sim \theta_k^{\mathrm{const}}.
  \label{I-theta-sh}
\end{equation}

\subsection{Post-Inflation Epoch with Dynamical~$f$}

When the decay constant starts to evolve at $a = a_\mathrm{i}$, the
fluctuations 
can get enhanced even on super-horizon scales. Let us suppose that $f$
shrinks with a power index
\begin{equation}
 n > \frac{5}{4},
\end{equation}
so that $\nu > 1$.
We compute the phase velocity fluctuations for super-horizon modes under
the dynamical~$f$ by connecting (\ref{I-dthetada-sh}) to (\ref{case-ab}),
yielding 
\begin{equation}
 \frac{d\theta_k}{d \ln a} \sim
  -\theta_k^{\mathrm{const}} \left( \frac{k}{a_\mathrm{i} H_\mathrm{i}}
			     \right)^2 
  \left( \frac{a}{a_\mathrm{i}} \right)^{2 n -\frac{3}{2}}
  =   -\theta_k^{\mathrm{const}}
     \left( \frac{k}{a H} \right)^2
  \left( \frac{f_{\mathrm{inf}}}{f} \right)^{2-\frac{5}{ 2 n}}.
  \label{II-dthetada-sh}
\end{equation}
The phase fluctuation is a sum of the constant mode and the growing
mode obtained by integrating the phase velocity,
\begin{equation}
 \theta_k \sim \theta_k^{\mathrm{const}}
  \left\{
   1 - \left(2 n - \frac{3}{2}  \right)^{-1}
   \left( \frac{k}{a H} \right)^2
  \left( \frac{f_{\mathrm{inf}}}{f} \right)^{2-\frac{5}{ 2 n}}
  \right\},
  \label{II-theta-sh}
\end{equation}
which shows that the growing mode dominates over the constant mode on wave
numbers~$k$ larger than 
\begin{equation}
 k_{\mathrm{grow}}(\tau) \equiv \left( 2 n - \frac{3}{2} \right)^{1/2}
   a H \left( \frac{f}{f_{\mathrm{inf}}} \right)^{1-\frac{5}{4 n}}.
\end{equation}
One sees that $k_{\mathrm{grow}}$ initially coincides with the horizon
scale, i.e. $k_{\mathrm{grow}}(\tau_\mathrm{i}) \sim a_\mathrm{i}
H_\mathrm{i}$, 
then $k_{\mathrm{grow}}$ becomes smaller than $a H$ as $f$ shrinks.
Thus the wave number $k_{\mathrm{grow}} (\tau_\mathrm{f})$ 
corresponds to the smallest $k$-mode (largest length scale) where the
axion fluctuation is enhanced due to the shrinking of~$f$. 

After $f$ becomes constant, i.e. $ a > a_\mathrm{f}$, the fluctuations
outside the horizon becomes constant once again.\footnote{When
naively connecting the super-horizon scalings (\ref{case-ab}) and
(\ref{case-cd}) across $a = a_\mathrm{f}$, the super-horizon
fluctuations might seem to grow even after $f$ has stopped its
evolution.
However it should be kept in mind that 
(\ref{case-ab}) and (\ref{case-cd}) only describe the asymptotic
behaviors in each epoch, and so does not necessarily provide good
approximations right after the transition.
To obtain the correct behavior of $\theta_k \sim \mathrm{const.}$,
one needs to take into account both terms in (\ref{Besseluto0}) and/or
the sub-leading terms that are dropped there.}

Let us focus on the wave number
that re-enters the horizon
when $f$ approaches its present value~$f_0$, i.e.,
\begin{equation}
 k_\mathrm{f} \equiv a_\mathrm{f} H_\mathrm{f}.
\end{equation}
The fluctuation on~$k_\mathrm{f}$ when $a = a_\mathrm{f}$ is
computed by 
extrapolating the super-horizon growing mode in~(\ref{II-theta-sh}) to
the time 
of horizon entry; in terms of the power spectrum~(\ref{Poftheta}) we find
\begin{equation}
 \mathcal{P}_\theta^{1/2} (\tau_\mathrm{f}, k_\mathrm{f})
 \sim  
\frac{k_\mathrm{f}^{3/2} \abs{\theta_{k_\mathrm{f}}^{\mathrm{const}}}}{
2^{1/2} \, \pi}
  \left( \frac{f_{\mathrm{inf}}}{f_0} \right)^{2-\frac{5}{ 2 n}}
  = \frac{H_{\mathrm{inf}}}{2  \pi f_0 }
  \left( \frac{f_{\mathrm{inf}}}{f_0} \right)^{1 - \frac{5}{2 n}},
  \label{thetakfaf}
\end{equation}
where we have ignored the factor
$(2n - \frac{3}{2})^{-1}$
in (\ref{II-theta-sh}).
Note that this expression does not depend directly on 
$H_{\mathrm{f}}$~itself; the enhanced fluctuation amplitude is independent of
when $f$ varies.

After $f$ approaches~$f_0$, i.e. $ a > a_\mathrm{f}$,
the wave mode~$k_\mathrm{f}$ is inside the
horizon and thus the fluctuation is damped as
\begin{equation}
 \mathcal{P}_\theta^{1/2} (\tau , k_\mathrm{f})
 \sim  
 \frac{H_{\mathrm{inf}}}{2  \pi f_0 }
 \left( \frac{f_{\mathrm{inf}}}{f_0} \right)^{1 - \frac{5}{2 n}}
 \frac{a_\mathrm{f}}{a}
 \qquad
 \mathrm{for}
 \quad
 \tau > \tau_\mathrm{f} .
  \label{4.3final}
\end{equation}
One can also check that in this epoch, until the
mode~$k_{\mathrm{grow}}(\tau_\mathrm{f})$ 
enters the horizon, the fluctuation spectrum has a plateau in the range
$aH \lesssim k \lesssim k_{\mathrm{f}}$, as we will see in Figure~\ref{fig:spectra}.  
Hence, denoting the time when $k_{\mathrm{grow}}(\tau_\mathrm{f})$
enters the horizon as $\tau = \tau_{\mathrm{grow}} $, the fluctuation amplitudes of the
modes coming into the horizon $k = a H$ are the same as~(\ref{4.3final}),
\begin{equation}
 \mathcal{P}_\theta^{1/2} (\tau, a H)
 \sim  
 \frac{H_{\mathrm{inf}}}{2  \pi f_0 }
 \left( \frac{f_{\mathrm{inf}}}{f_0} \right)^{1 - \frac{5}{2 n}}
 \frac{a_\mathrm{f}}{a}
 \qquad
 \mathrm{for}
 \quad
 \tau_{\mathrm{f}} < \tau < \tau_{\mathrm{grow}}.
\label{kaH3}
\end{equation}

\subsection{Backreaction}

Before closing this section, let us comment on the backreaction from the
phase fluctuations on the expanding universe.
By computing the energy momentum tensor sourced by the axion kinetic term
in the action~(\ref{theta-action}),
the energy density of the phase fluctuations is obtained as
\begin{equation}
 \rho_\theta =
\langle -\tensor{T}{^\theta_0^0} (\tau,
 \boldsymbol{x})  \rangle = 
\int \frac{dk}{k}\, 
  \frac{k^3 H^2 f^2 }{4 \pi^2}
  \left(
   \left| \frac{d \theta_k}{d \ln a} \right|^2 +
   \frac{k^2 \abs{\theta_k}^2}{a^2 H^2} 
	  \right).
  \label{a-density}
\end{equation}
In the above discussions we have treated the decay constant as a classical
background that, while decreasing in time, injects energy into the phase
fluctuations. Here, since the energy density of the field(s) that sets the
time evolution of the decay constant cannot exceed the total energy
density of the universe, 
$\rho_\theta \ll 3 M_p^2 H^2$ should always be satisfied.
In other words, the phase fluctuations cannot be enhanced to become
$\rho_\theta \sim 3 M_p^2 H^2$ without significantly backreacting on the field(s)
driving the $f$-dynamics, and also on the background universe. 
The coupling between $f$ and $\theta$ in the axion kinetic term further
sources direct backreaction from the enhanced axion fluctuations.
We also note that if $\rho_\theta$ is enhanced to become comparable to
$\sim f^4$, then it can backreact on the radial component of the PQ scalar.
On the one hand, the backreaction may force the PQ scalar to climb up
the Mexican hat potential, while on the other hand the enhanced
angular velocity may prevent the PQ scalar from approaching the origin.
It would be interesting to study whether the enhanced axion fluctuations
assist or prevent the restoration of the PQ symmetry in the
post-inflation universe.

\section{Cosmological Implications of Axion Perturbations}
\label{sec:cosmo}

Let us now discuss the cosmological implications of the enhanced axion
fluctuations.
We will show that while a shrinking~$f$ may relax constraints from
large-scale isocurvature perturbations,
the strongly enhanced axion fluctuations on small scales can lead to
disastrous formation of axionic domain walls.

\subsection{Domain Walls}

In the previous section we have seen that, after the decay constant
approaches its present-day value,
the enhanced axion
fluctuation~$\mathcal{P}_\theta^{1/2}$ holds a plateau 
over the wave modes $ a H \lesssim k \lesssim k_{\mathrm{f}}$.
These enhanced fluctuations that are scale-invariant up to the Hubble radius
damps as $\propto a^{-1}$ as the universe expands. 
Eventually the universe undergoes reheating,
and the axion's effective potential (\ref{theta-action}) emerges when
the universe cools down to $T \sim \Lambda_{\mathrm{QCD}}$;
if the fluctuation $\mathcal{P}_\theta^{1/2}$ then 
is still larger than the period of the potential
\begin{equation}
 \Delta \theta = \frac{2 \pi }{N},
  \label{a-period}
\end{equation}
then the axion settles down to different potential minima in different
patches of the universe, leading to the formation of cosmic domain
walls.

Here a few things are worth noting.
Since the PQ symmetry breaking happens before inflation in our case of
interest, unlike the standard axionic domain
walls~\cite{Vilenkin:1982ks,Sikivie:1982qv,Linde:1990yj,Lyth:1992tx},
the walls produced here are {\it not} connected to cosmic strings.
(To be more precise, some walls are bounded by strings produced before inflation, but the number
density of the strings is extremely tiny because of the dilution
during inflation.)
The axion potential~(\ref{theta-action}) can be viewed as a
one-dimensional  
potential with an infinite number of degenerate vacua,
and we stress that the formation of domain walls is due to the axion
fluctuations being spread over multiple vacua,\footnote{Works such
as~\cite{Kitajima:2014xla,Daido:2015bva,Daido:2015cba,Hebecker:2016vbl} have also discussed domain wall
formation due to multiple vacua lying within the field fluctuation.}
with the number of different vacua being
$\sim \mathcal{P}_\theta^{1/2} / \Delta \theta $.
This is why domain walls can form
without having wall junctions, 
even with numerous vacua. 
It should also be noted that in the one-dimensional axion
potential~(\ref{theta-action}), each vacuum only has two vacua on its sides.
As a consequence, adjacent domains in the universe that are
separated by a single wall can only hold vacua that are adjacent also
in the axion potential.

The important question for cosmology is whether infinitely large domain
walls form. If they do, they would overclose the universe and spoil the
cosmological expansion history.
When $ \mathcal{P}_\theta^{1/2} / \Delta \theta \gg 1$,
with the vast number of vacua,
naively, it would seem difficult for a single wall to extend throughout
the universe.
However, this is not necessarily the case due to the
specific features mentioned above.
In order to see this, let us refer to the unperturbed background field
value of the axion before 
the potential emerges by~$\bar{\theta}$,
and the field value that becomes the nearby potential minimum 
by~$\theta_{\mathrm{min}\, s}$.
In other words, $\bar \theta$ and $\theta_{\mathrm{min}\, s}$ satisfy
$\theta_{\mathrm{min}\, s}- \Delta\theta/2 < \bar \theta < \theta_{\mathrm{min}\, s} + \Delta\theta/2$.
We also label the other vacua along the one-dimensional axion potential as
\begin{equation}
 \theta_{\mathrm{min}\, s\pm 1} = \theta_{\mathrm{min}\, s} \pm \Delta
  \theta,
  \quad
   \theta_{\mathrm{min}\, s\pm 2} = \theta_{\mathrm{min}\, s} \pm 2 \Delta
   \theta,
   \quad \cdots,
\end{equation}
and further divide the vacua into two groups, depending on which side of 
$\theta_{\mathrm{min}\, s}$ they are located on,
\begin{equation}
 \Theta_{\mathrm{min} -} = \{
   \cdots, \,
   \theta_{\mathrm{min}\, s-1}, \,
   \theta_{\mathrm{min}\, s}
  \},
\quad
 \Theta_{\mathrm{min} +} = \{
\theta_{\mathrm{min}\, s + 1}, \,
  \theta_{\mathrm{min}\, s + 2}, \,  \cdots \,
\}.
\end{equation}
(Here we included $\theta_{\mathrm{min}\, s}$ in $\Theta_{\mathrm{min}
-}$, but one may choose to include it in the other group as well.)
Now, let us crudely model the universe as a collection of cells with size of
the Hubble radius, each of which
holding a vacuum belonging to either group with an 
independent probability.
If $ \mathcal{P}_\theta^{1/2} / \Delta \theta
\gg 1$ when the potential emerges,
then since the axion fluctuation spreads equally on both sides of~$\bar{\theta}$,
the probability for each cell to fall in 
$\Theta_{\mathrm{min} -}$ or $\Theta_{\mathrm{min} +}$ 
would be roughly the same, $p \sim 0.5$.
Here, in order to tell whether the collection of cells with 
$\Theta_{\mathrm{min}-}$ or $\Theta_{\mathrm{min}+}$ forms an infinitely
large connected region,
useful insights can be obtained from studies of percolation theory:
Since the percolation threshold~$p_\mathrm{c}$ in three dimensions is
typically smaller than 0.5
(e.g. $p_c \sim 0.3$ for a cubic lattice~\cite{Stauffer:1978kr}),
an infinitely large region with vacua belonging to 
$\Theta_{\mathrm{min}-}$ exists, and likewise for
$\Theta_{\mathrm{min}+}$.
This means that an infinitely large wall between
$\Theta_{\mathrm{min}-}$ and $\Theta_{\mathrm{min}+}$ appears in the
universe. 
Because of the specific vacuum distribution discussed above,
this indicates that there exists an infinitely large domain wall between the 
$\theta_{\mathrm{min}\, s}$ and $\theta_{\mathrm{min}\, s+1}$~vacua.
Such infinitely large domain walls can also form between other pairs of
vacua as well. 
Over time the domain walls can annihilate each other,
however, as each patch of the universe randomly holds
$\Theta_{\mathrm{min}-}$ or $\Theta_{\mathrm{min}+}$ at $T \sim
\Lambda_{\mathrm{QCD}}$, we expect there to be always at least one
domain wall that extends throughout the observable universe. 
The domain walls without junctions will likely obey the usual scaling
law~\cite{Press:1989yh,Garagounis:2002kt,Leite:2011sc} soon after formation.

Therefore we conclude that axionic domain walls would form and overclose
the universe, if 
\begin{equation}
 \left. \mathcal{P}_\theta^{1/2} \right|_{T \sim \Lambda_{\mathrm{QCD}}}
  \gtrsim \Delta \theta
\label{eq5.4}
\end{equation}
is satisfied on the mode~$k_{\mathrm{f}}$,
and thus also for the horizon size mode,
when the axion potential emerges.
A full treatment of domain walls from enhanced axion fluctuations
is beyond the scope 
of this paper, but it would be 
interesting to carry out detailed analyses (which presumably
requires numerical simulations) of the formation and evolution of the
wall network.

\subsection{Isocurvature Perturbations}

When the axion constitutes a non-negligible fraction of CDM, the
large-scale fluctuations of the axion serve as CDM isocurvature
perturbations, which are severely constrained by CMB measurements. 
Here we briefly review how such isocurvature constraints on the axion
are relaxed with a dynamical decay
constant~\cite{Linde:1990yj,Linde:1991km}.

Let us focus on the vicinity of one of the
minima~$\theta_{\mathrm{min}}$ of the axion 
potential~(\ref{theta-action}),
\begin{equation}
 N^2 (\theta - \theta_{\mathrm{min}})^2 \lesssim 1,
\end{equation}
where the potential is approximately quadratic,
\begin{equation}
 V \propto (\theta - \theta_{\mathrm{min}})^2.
  \label{quad-pot}
\end{equation}
Then the energy density (isocurvature) fluctuation of the axion is estimated as
\begin{equation}
 \frac{\delta \rho_\theta }{\rho_\theta  }
  \sim   \frac{2 \delta \theta }{
  \bar{\theta} - \theta_{\mathrm{min}}
  }, 
\label{eq5.7}
\end{equation}
where $\bar{\theta}$ denotes the axion's unperturbed background field
value before the axion acquires its potential.\footnote{As mode-mode
couplings are absent before the axion potential emerges, 
the unperturbed value~$\bar{\theta}$ is unaffected by the enhanced
small-scale fluctuations~$\theta_k$.
However the large- and small-scale modes can interact after the
potential emerges;
in the extreme case of~(\ref{eq5.4}), domain walls are formed and even
the background value~$\bar{\theta}$ is affected by the small-scale
fluctuations.
In such cases the enhanced fluctuations are preserved by the walls and thus
one cannot expand around~$\bar{\theta}$ as in~(\ref{eq5.7}).
We also note that if $\bar{\theta}$ is initially close to the potential hilltop
(as in cases where anharmonic effects are important), the
criteria~(\ref{eq5.4}) for domain wall formation is relaxed.}
Hence the power spectrum of the CDM isocurvature perturbations, which we denote
by~$\mathcal{P}_S$, is expressed in terms of the field
fluctuation spectrum~(\ref{Poftheta}) as
\begin{equation}
 \mathcal{P}_S (k) \sim
\frac{4 \mathcal{P}_\theta (k)
  }{(\bar{\theta} - \theta_{\mathrm{min}})^2}
\left( \frac{\Omega_\theta }{\Omega_{\mathrm{CDM}}} \right)^2
    \sim
4 N^2 \mathcal{P}_\theta (k)
\left( \frac{f_0 / N}{10^{12}\, \mathrm{GeV}} \right)^{7/6}
 \frac{\Omega_\theta }{\Omega_{\mathrm{CDM}}} .
\label{PofS}
\end{equation}
Here $\Omega_{\theta (\mathrm{CDM})}$ is the axion (CDM) abundance today,
and upon moving to the far right hand side we have used the relation
between the axion abundance and the phase~\cite{Turner:1985si}
(note that, as we are considering the axion potential to emerge after
the decay constant has approached its present-day value,
the dynamics of the 
unperturbed axion is basically unmodified
unless axionic walls are formed):
\begin{equation}
 \frac{\Omega_\theta }{\Omega_{\mathrm{CDM}}} \sim
   N^2 (\bar{\theta} -   \theta_{\mathrm{min}})^2 
   \left( \frac{f_0 / N}{10^{12}\, \mathrm{GeV}}   \right)^{7/6}.
   \label{a-abundance}
\end{equation}
A scale-invariant and uncorrelated CDM isocurvature is
constrained on CMB scales by {\it Planck}~\cite{Ade:2015lrj} as
\begin{equation}
 \mathcal{P}_S (k_*)
  \lesssim   0.040 \times 
  \mathcal{P}_{\zeta} (k_*)
  \quad
  (95\%\, \mathrm{C.L.}, \, \, \mathrm{TT,TE,EE+lowP}  )
 \label{isocon}
\end{equation}
where the pivot scale is $k_*/ a_0 = 0.05 \,
\mathrm{Mpc}^{-1}$,
and the adiabatic power is 
$ \mathcal{P}_{\zeta} (k_*) \approx 2.2 \times 10^{-9} $.

Supposing $k_* \ll k_{\mathrm{grow}} (\tau_\mathrm{f})$
so that the axion fluctuations on the CMB scales are not affected by the
evolution of~$f$, 
then from (\ref{inf-theta-sh}),
\begin{equation}
 \mathcal{P}_\theta (k_*) \sim
  \left( \frac{H_{\mathrm{inf}}}{2 \pi f_{\mathrm{inf}}}  \right)^2.
\end{equation}
This combined with (\ref{PofS}) and
(\ref{isocon}) gives an upper bound on the inflation scale,
\begin{equation}
  H_{\mathrm{inf}}
   \lesssim
   10^7\, \mathrm{GeV} \; 
    \frac{f_{\mathrm{inf}}}{f_0}
    \left(\frac{f_0/N}{10^{12}\, \mathrm{GeV}}\right)^{5/12}
 \left( \frac{\Omega_{\mathrm{CDM}}}{\Omega_\theta } \right)^{1/2}.
 \label{HinfUB}
\end{equation}
For instance if $f_0 / N = 10^{12}\, \mathrm{GeV}$ and $\Omega_\theta =
\Omega_{\mathrm{CDM}}$
(which is realized with
$N \abs{\bar{\theta} - \theta_{\mathrm{min}}} \sim 1$,
cf.~(\ref{a-abundance})),
without the evolution of~$f$ (i.e. $f_{\rm inf} = f_0$), 
the CMB bound on isocurvature perturbations requires the
inflation scale to be as low as $H_{\mathrm{inf}} \lesssim 10^7\, \mathrm{GeV}$.

On the other hand when the decay constant is allowed to shrink,
then the upper
bound on~$H_{\mathrm{inf}}$ is relaxed by a factor of $f_{\mathrm{inf}}
/ f_0$.
However with a dynamical~$f$,
the axion perturbations can get enhanced on smaller scales, as we
have seen in the previous sections.
We will study this effect in more detail in the following subsection.

Before moving on, we should also remark that the above
estimation of the isocurvature perturbations breaks down
for $N \abs{\bar{\theta} - \theta_{\mathrm{min}}} \gtrsim 1$,
i.e., when the axion is initially sitting away from the 
potential minimum.
In particular for cases with $\Omega_\theta = \Omega_{\mathrm{CDM}}$,
such anharmonic effects become important for $f_0/ N \lesssim 10^{11}\,
\mathrm{GeV}$;
there the expression~(\ref{a-abundance}) for the axion abundance
is modified, and most importantly,
the isocurvature perturbations become much larger than in~(\ref{PofS}).
Consequently, the upper bound on the inflation scale
is much stronger than in~(\ref{HinfUB}).
See \cite{Lyth:1991ub,Kobayashi:2013nva} for detailed discussions on the
anharmonic enhancement of the axion isocurvature perturbations.

\subsection{Enhancement of Small-Scale Perturbations}

We now study two example
cases and see how the small-scale axion fluctuations are actually
enhanced, possibly leading to formation of domain walls.
In both examples
we assume the time evolution of the decay constant as was discussed in
Section~\ref{sec:DDC},
and further suppose the axion to account for all the dark matter,
\begin{equation}
 \Omega_\theta = \Omega_{\mathrm{CDM}}. 
\end{equation}

\subsubsection{Case 1}

We first study the case with
\begin{equation}
 \frac{f_0}{N} = 10^{11}\, \mathrm{GeV},
  \quad
 H_{\mathrm{inf}} = 10^{14}\, \mathrm{GeV},
\end{equation}
where the inflation scale is set to be roughly the current upper
limit~\cite{Ade:2015lrj}.
Were it not for a dynamical~$f$, with these parameters
the PQ symmetry breaking would happen after inflation and then
domain walls would overclose the universe unless $N = 1$.
However this can be avoided with a time-dependent~$f$;
one sees from (\ref{HinfUB}) that even the isocurvature
constraint can be satisfied if $f$ shrinks by
\begin{equation}
 \frac{f_0}{f_{\mathrm{inf}}} \lesssim 10^{-7},
  \label{5.15}
\end{equation}
i.e. $f_{\mathrm{inf}} / N \gtrsim 10^{18}\, \mathrm{GeV}$.\footnote{The
axion decay constant during inflation is required to be comparable to 
or larger than the Planck scale, in order to be consistent with high-scale inflation 
that saturates the current upper bound on the tensor-to-scalar ratio~\cite{Higaki:2014ooa}.}

Let us now look into the small-scale fluctuations. 
The axion fluctuation on the mode~$k_{\mathrm{f}}$
when the evolution of $f$ ceases 
is estimated from (\ref{thetakfaf}) as
\begin{equation}
 \frac{\mathcal{P}_\theta^{1/2} (\tau_\mathrm{f}, k_\mathrm{f})}{\Delta
  \theta }
  \sim \frac{10^3}{(2 \pi)^2} \times
  \left(\frac{f_{\mathrm{inf}}}{f_0}\right)^{1 - \frac{5}{2 n}},
\label{5.16}
\end{equation}
in terms of the potential period~(\ref{a-period}).
In particular when $n > 5/2$,
the constraint~(\ref{5.15}) gives a lower bound on the fluctuation.
For instance, 
\begin{equation}
 \mathrm{for} \quad n = 4,
    \qquad
 \frac{\mathcal{P}_\theta^{1/2} (\tau_\mathrm{f}, k_\mathrm{f})}{\Delta
  \theta }
  \gtrsim 10^4.
  \label{case1-n4}
\end{equation}
After $\tau = \tau_\mathrm{f}$, the universe eventually undergoes
reheating (as we have been assuming that $f$ evolves during the
matter-dominated era), and the axion potential emerges when $ T \sim
\Lambda_{\mathrm{QCD}}$. 
Since the fluctuations on the mode~$k_\mathrm{f}$ as well as the horizon
size mode are damped as $\propto a^{-1}$ 
from the value~(\ref{case1-n4}), 
one sees that domain walls would form if the axion potential emerges
before the universe expands by $a / a_{\mathrm{f}} \sim 10^4$ since when
$f$ has approached its present-day value.

We also note that when $n$ is as large as $n \gtrsim 12$, then
the energy density of the fluctuations~(\ref{a-density}) becomes at least
comparable to the total density of the 
universe at $\tau = \tau_{\mathrm{f}}$, which indicates the breakdown of our
treatment of the decay constant and the cosmological expansion as homogeneous
backgrounds.
With such a large~$n$, the backreaction from the axion fluctuations 
becomes non-negligible 
before the time $\tau = \tau_{\mathrm{f}}$.
On the other hand, whether the fluctuation energy density exceeds~$f^4$
depends also on when $f$ varies.
For instance if $f$ shrinks quickly soon after inflation,
then $f^4$ can become smaller than the total density of the universe.
In such cases one will have to worry about the recovery of the PQ
symmetry due to the gravitational background or the enhanced axion
fluctuations. 

\vspace{\baselineskip}

\begin{figure}[t!]
 \begin{minipage}{.48\linewidth}
  \begin{center}
 \includegraphics[width=\linewidth]{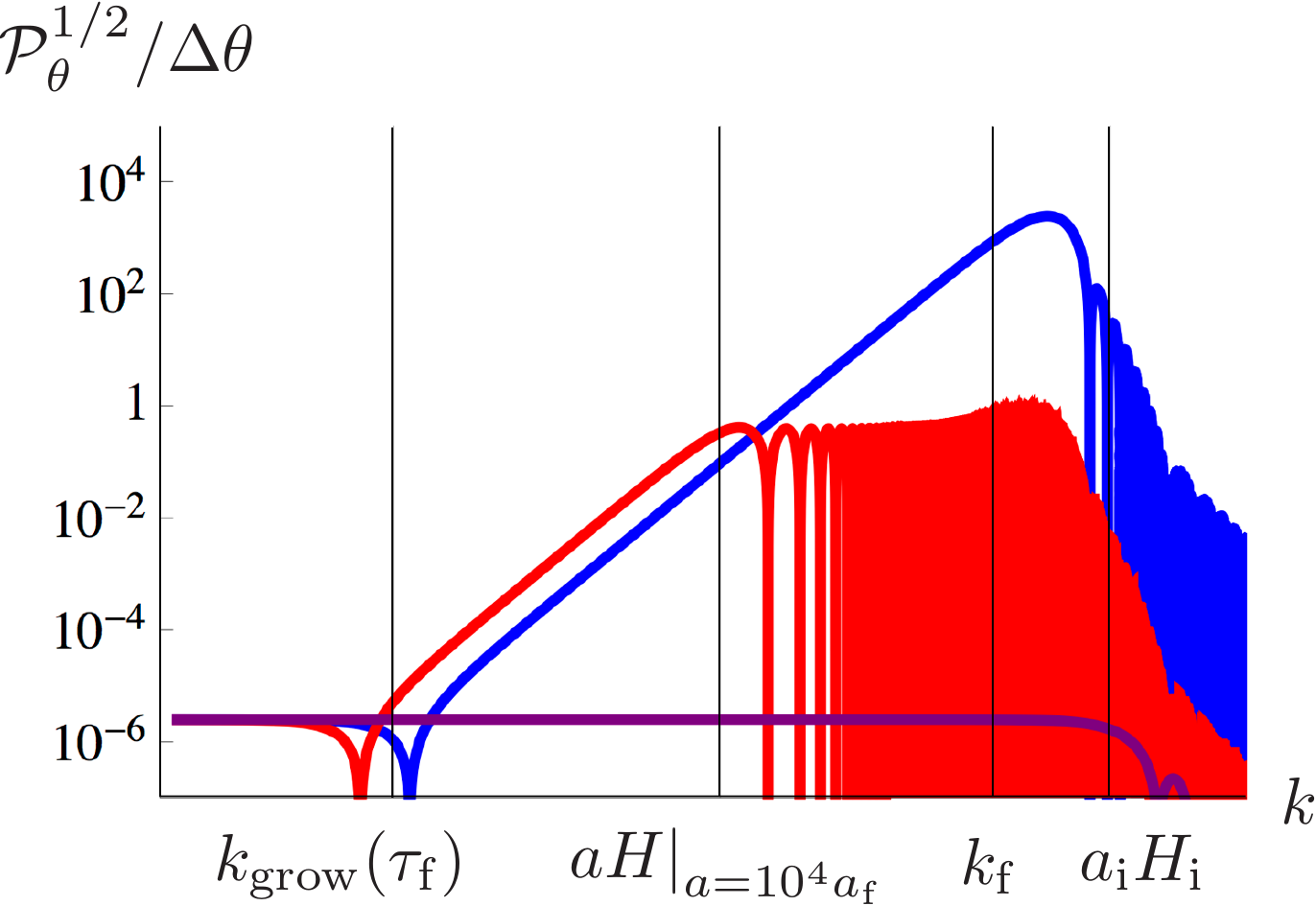}
  \end{center}
 \end{minipage} 
 \begin{minipage}{0.01\linewidth} 
  \begin{center}
  \end{center}
 \end{minipage} 
 \begin{minipage}{.48\linewidth}
  \begin{center}
 \includegraphics[width=\linewidth]{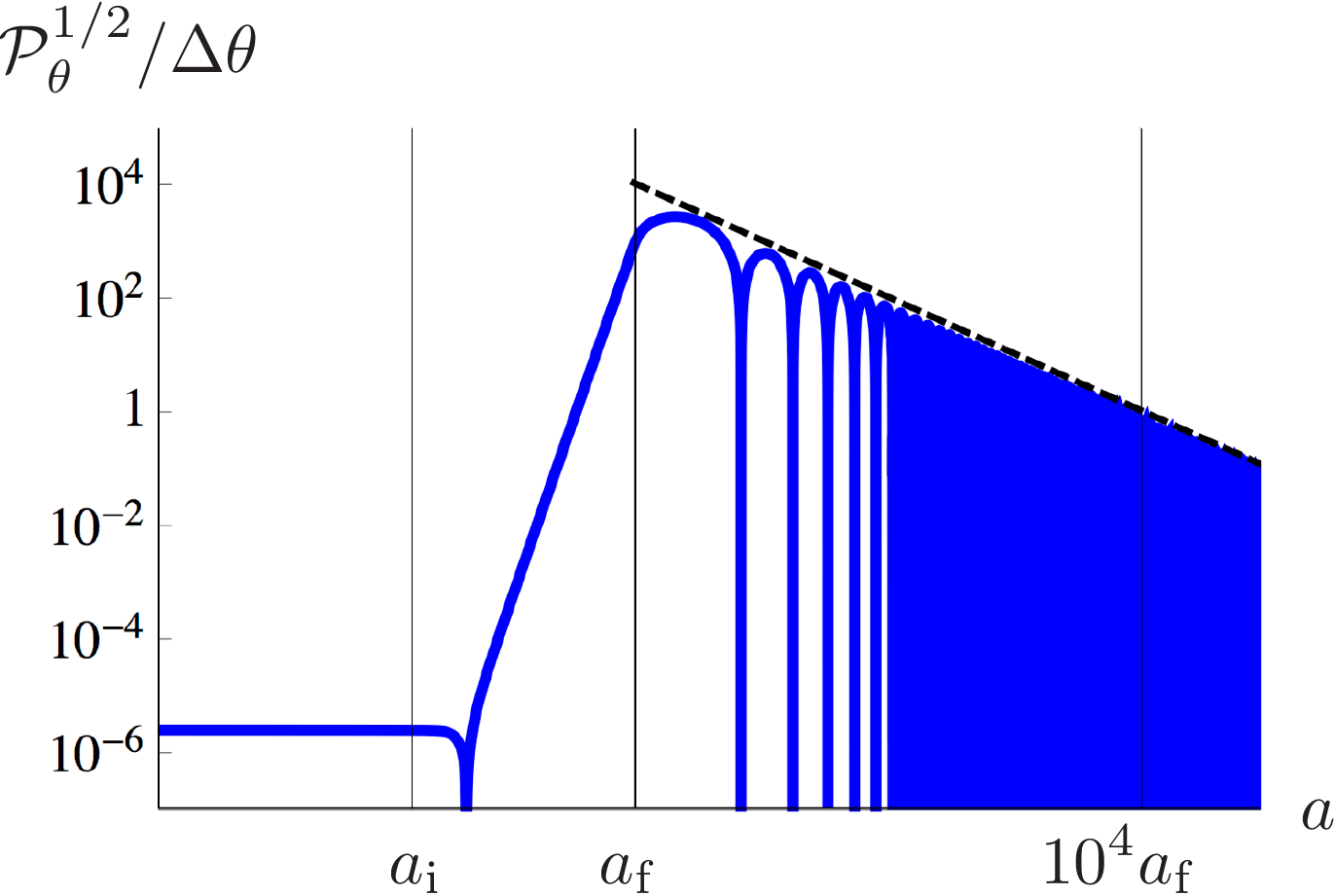}
  \end{center}
 \end{minipage} 
 \caption{Axion phase fluctuation~$\mathcal{P}_\theta^{1/2}$ in
 terms of the potential period~$\Delta \theta$, for the case of
 $H_{\mathrm{inf}} = 10^{14}\, \mathrm{GeV}$, $f_0 / N = 10^{11}\,
 \mathrm{GeV}$, $f_{\mathrm{inf}} / N = 10^{18}\, \mathrm{GeV}$, and $n
 = 4$. 
 Left: Fluctuation spectrum as a function of wave number~$k$, when 
 $a = a_{\mathrm{i}}$ (purple),  $a_{\mathrm{f}}$ (blue), $10^4 a_{\mathrm{f}}$ (red). 
 Right: Time-dependence of fluctuation on wave
 number~$k_{\mathrm{f}}$. The dashed line shows the estimate obtained
 from~(\ref{5.16}). In this case, axionic domain walls would form if the
 axion potential emerges before $ a \sim 10^4 a_{\mathrm{f}}$.} 
 \label{fig:spectra}
\end{figure}

The fluctuation $\mathcal{P}_\theta^{1/2} / \Delta \theta $ is shown in
Figure~\ref{fig:spectra}, for 
the case of $n = 4$ and $f_{\mathrm{inf}} / N = 10^{18}\,
\mathrm{GeV}$.
Here we have plotted the exact solution for~$\theta_k$, obtained by
starting from the initial conditions (\ref{4.2}) and (\ref{4.3}) set
during inflation, and then connecting to the general solutions (\ref{generalsol}) and
(\ref{gendthetada}) in each epoch of
the $f$~evolution~(\ref{f-evo}) after inflation.
Here it should be noted that the fluctuation amplitude at
$\tau = \tau_\mathrm{f}$~(\ref{thetakfaf}) depends on the
ratio~$f_{\mathrm{inf}} / f_0$, but not directly on 
when $f$ evolves;
hence for the wave modes shown in the figures, the shape of the
fluctuation spectrum is basically independent of
parameters such as~$H_{\mathrm{f}}$,
as long as $H_{\mathrm{inf}}$,
$f_{\mathrm{inf}} / N$, $f_{\mathrm{0}} / N$, and $n$ are fixed.
The left figure displays the fluctuation spectrum as a function of~$k$
when $a = a_{\mathrm{i}}$ (purple), 
$a_{\mathrm{f}}$ (blue), $10^4 a_{\mathrm{f}}$ (red).
One sees that the initially flat super-horizon spectrum is 
enhanced while $f$ evolves on wave numbers 
$k \gtrsim k_{\mathrm{grow}}(\tau_\mathrm{f})$.
After $f$ ceases to evolve, the spectrum is flattened as the fluctuations
enter the horizon and are damped.
Hence, until the mode~$k_{\mathrm{grow}}(\tau_\mathrm{f})$ enters the
horizon,
the fluctuations coming into the horizon have roughly the same amplitude as 
that on the mode~$k_{\mathrm{f}}$ inside the horizon,
cf.~(\ref{kaH3}).

The right figure shows the fluctuation as a function of~$a$, focusing on
the wave number~$k_{\mathrm{f}}$. 
One sees that the fluctuation's dynamical mode dominates over the
constant mode some time after $f$ starts to evolve,
cf.~(\ref{II-theta-sh}).
For $a \geq a_{\mathrm{f}}$,
the fluctuation amplitude is analytically estimated by multiplying 
(\ref{5.16}) by the damping factor $a_{\mathrm{f}} / a$,
which is shown as the dashed line in the figure.
This estimate is seen to agree well with the oscillation amplitude of the
exact solution.
In order to avoid domain walls from forming in this case,
the axion potential should not emerge 
at least until $a \sim 10^4 a_{\mathrm{f}}$,
before which $\mathcal{P}_\theta^{1/2} (k = a H)$ is larger
than~$\Delta \theta$.

\subsubsection{Case 2}

Now let us consider the parameter set 
\begin{equation}
 \frac{f_0}{N} = 10^{10}\, \mathrm{GeV}, \quad
  H_{\mathrm{inf}} = 10^{14}\, \mathrm{GeV}.
  \label{5.19}
\end{equation}
For this~$f_0/N$,
the unperturbed axion field value needs to be initially located close to
the potential hilltop in order for the axion to account for all the dark
matter, 
as then the onset of the axion oscillation 
is delayed and the axion abundance is enhanced compared to that
shown in~(\ref{a-abundance}).
However, the anharmonic effects also strongly enhance the axion
isocurvature fluctuations, resulting in a much stronger upper bound on
the inflation scale than shown in~(\ref{HinfUB}).
For $f_0 / N = 10^{10}\, \mathrm{GeV}$ and $\Omega_\theta =
\Omega_{\mathrm{CDM}}$, were it not for a dynamical~$f$, 
the upper bound on $H_{\mathrm{inf}}$ is actually
$\sim 10^2\, \mathrm{GeV}$; see Figure~3 of~\cite{Kobayashi:2013nva}. 
On the other hand when $f$ is allowed to vary, the isocurvature bound constrains 
the rescaled inflation scale~$(f_0 / f_{\mathrm{inf}})
H_{\mathrm{inf}}$,\footnote{This is clearly seen in the
bound~(\ref{HinfUB});
after multiplying both sides by $f_0 / f_{\mathrm{inf}}$, the expression
can be read as an upper bound on~$(f_0 /
f_{\mathrm{inf}})H_{\mathrm{inf}} $.}
and thus we have 
$(f_0 / f_{\mathrm{inf}}) H_{\mathrm{inf}} \lesssim 10^2\, \mathrm{GeV}$.
Hence for the parameters~(\ref{5.19}),
the isocurvature constraint is satisfied if $f$ shrinks as
\begin{equation}
 \frac{f_0}{f_{\mathrm{inf}}} \lesssim 10^{-12},
\end{equation}
i.e. $f_{\mathrm{inf}} / N \gtrsim 10^{23}\, \mathrm{GeV}$.
Although such a large decay constant may be difficult to obtain in a
controlled setting~\cite{Banks:2003sx,Svrcek:2006yi}, here let us proceed
with this bound.

The large hierarchy between $f_{\mathrm{inf}}$ and
$f_0$ significantly enhances the small-scale fluctuations. 
For the mode~$k_{\mathrm{f}}$, cf.~(\ref{thetakfaf}), the fluctuation is
\begin{equation}
 \frac{\mathcal{P}_\theta^{1/2} (\tau_\mathrm{f}, k_\mathrm{f})}{\Delta
  \theta }
  \sim \frac{10^4}{(2 \pi)^2} \times
  \left(\frac{f_{\mathrm{inf}}}{f_0}\right)^{1 - \frac{5}{2 n}}.
\end{equation}
Hence, e.g.,
\begin{equation}
 \mathrm{for} \quad n = 4,
    \qquad
 \frac{\mathcal{P}_\theta^{1/2} (\tau_\mathrm{f}, k_\mathrm{f})}{\Delta
  \theta }
  \gtrsim 10^7
\end{equation}
and domain walls form if the axion potential emerges
before the universe expands by $a / a_{\mathrm{f}} \sim 10^7$ after the
evolution of $f$ ceases. 
For $n \gtrsim 5$, the $k_{\mathrm{f}}$~mode contribution to the
fluctuation density~(\ref{a-density}) becomes at least
comparable to the total density of the 
universe at $\tau = \tau_{\mathrm{f}}$,
and thus the backreaction from the fluctuations needs to be taken into
account.

\section{Conclusions}
\label{sec:conc}

While a dynamical decay constant of the QCD axion
can help accommodate high-scale inflation,
we have shown that 
it also enhances the small-scale axion isocurvature perturbations.
This effect stretches out to super-horizon modes,
and thus the enhanced perturbation modes will
continue to enter the horizon for some time after the decay constant
stops its evolution.
The axion fluctuation can become much larger than the potential
period,
thus unless the fluctuation redshifts away by the time
the universe cools down to $T \sim \Lambda_{\mathrm{QCD}}$, 
it would lead to the formation of axionic domain walls.
These walls produced by the enhanced axion fluctuations are not bounded by
cosmic strings, and thus would dominate the universe even in the case of
$N = 1$. 

In our analyses we have assumed the decay constant~$f$ to vary as a
power-law of the scale factor.
However in actual cases the dynamics of~$f$ could be more complicated.
If, for instance, $f$ instead oscillates around its final
value~$f_0$, then the axion perturbations may get a ``kick'' during the
first oscillation, and be enhanced much stronger than in the cases
analyzed in this paper. 
The oscillations of~$f$ could further induce resonant
amplifications of the sub-horizon axion perturbations~\cite{Tkachev:1995md,Kasuya:1996ns,Kasuya:1997ha,Kasuya:1998td,Tkachev:1998dc,Kearney:2016vqw}.
We also note that, for the evolution of $f$ to take place not too far
from the QCD phase transition, a relatively light degree of freedom is required.
In a simple case, the mass of the field that sets the axion decay
constant~$f$ needs to be much
lighter than~$f$ itself. Such a hierarchy in scales is realized
in a supersymmetric axion model, where the saxion (corresponding to the
radial component~$r$, and thus setting~$f$) acquires a mass only from
supersymmetry breaking effects. When the fluctuations of the axion are enhanced
by the dynamical $f$, the saxion may also acquire large fluctuations, which results in
large spatial fluctuations of $f$.
It would be important to study the evolution of perturbations in
explicit models of axions with dynamical decay constants.

While we have mainly focused on the QCD axion in this paper,
similar discussions hold also for axion-like fields with dynamical
decay constants, or more generally,
for fields that have kinetic terms with time-dependent coefficients.
If such a field possesses a periodic potential 
at energy scales higher than the QCD scale, then
even if the evolution of the decay constant happens in the very early
times, the enhanced perturbations could still be transformed into domain
walls before redshifting away.
(However in such cases one may have to take into account the effect
of the potential, and in particular the induced mode-mode coupling,
during the evolution of the decay constant.) 
We also note that, depending on when the decay constant evolves, the
enhanced axion(-like) isocurvature modes may leave observable signals
for upcoming CMB and large scale structure surveys.

\section*{Acknowledgments}

T.K. would like to thank the Particle Theory and Cosmology Group of
Tohoku University for hospitality during the initiation of this work.
T.K. acknowledges support from the INFN INDARK PD51 grant.
This work is supported by MEXT KAKENHI Grant Numbers 15H05889 and 15K21733 (F.T.),
JSPS KAKENHI Grant Numbers  26247042(F.T.), and 26287039 (F.T.),
World Premier International Research Center Initiative (WPI Initiative), MEXT, Japan (F.T.).

%\newpage

%\appendix

%\section{Appendix A}
%\label{app:A}

%\clearpage

\end{document}